# Symmetry-breaking-induced multifunctionalities of two-dimensional chromium-based materials for nanoelectronics and clean energy conversion

Lei Li[1], Tao Huang[1], Kun Liang[1], Yuan Si[1], Ji-Chun Lian[1], Wei-Qing Huang[1*], Wangyu Hu[2], Gui-Fang Huang[1#]

[1] Department of Applied Physics, School of Physics and Electronics, Hunan University, Changsha 410082, China

[2] School of Materials Science and Engineering, Hunan University, Changsha 410082, China

**Abstract**: Structural symmetry-breaking that could lead to exotic physical properties plays a crucial role in determining the functions of a system, especially for two-dimensional (2D) materials. Here we demonstrate that multiple functionalities of 2D chromium-based materials could be achieved by breaking inversion symmetry via replacing Y atoms in one face of pristine CrY (Y=P, As, Sb) monolayers with N atoms, i.e., forming Janus $Cr_2NY$ monolayers. The functionalities include spin-gapless, very low work function, inducing carrier doping and catalytic activity, which are predominately ascribed to the large intrinsic dipole of Janus $Cr_2NY$ monolayers, making them having great potentials in various applications. Specifically, $Cr_2NSb$ is found to be a spin-gapless semiconductor, $Cr_2NP$ and $Cr_2NHPF$ could simultaneously induce *n*- and *p*-type carrier doping for two graphene sheets with different concentrations (forming intrinsic *p*-*n* vertical junction), and $Cr_2NY$ exhibits excellent electrocatalytic hydrogen evolution activity, even superior to benchmark Pt. The results confirm that breaking symmetry is a promising approach for the rational design of multifunctional 2D materials.

---

*. Corresponding author. *E-mail address:* wqhuang@hnu.edu.cn
#. Corresponding author. *E-mail address:* gfhuang@hnu.edu.cn

## I. Introduction

Symmetry principles that play an important role with respect to the laws of nature have been universally used in scientific community, and can simplify the solution of problems, thus easily to disclose the key conclusions. When breaking symmetry of a system, bizarre physical properties such as plasmonic exceptional points and valley Hall effect would appear [1-5]. The past few years have witnessed rapid development of symmetry-breaking two-dimensional (2D) materials. Notably, Janus transition-metal dichalcogenides (TMDs) has become a focus and frontier of 2D materials with mirror symmetry breaking, because their three-atom-layer structure naturally possesses a possibility to intrinsically break the out-of-plane mirror symmetry [6]. Experimentally, a semiconducting Janus MoSSe that its one face consists of Se atoms and the other is S atoms was synthesized in 2017 independently by Lu *et al*. [7] and by Zhang *et al*. [8]. This breakthrough has inspired a surge of theoretical and experimental research into Janus 2D materials, including but not limited to MoSSe, which exhibit unique physical and chemical properties different from their symmetrical counterparts. For instance, long exciton radiative recombination lifetime was found in Janus MoSSe and WSSe monolayers [9], considerable spin Hall conductivities [10], efficient water-splitting performance [11] and tunable electronic and electron transport properties [12] were predicted in Janus MoSSe. Due to their unique structure and properties, Janus 2D materials show great potential applications in nanoelectronics (such as valleytronics [13] and gas sensing [14]) and clean energy conversion (such as photovoltaics [15, 16] and water splitting [17, 18]).

Recently, 2D chromium-based materials have received great attention due to their unique structural topology and intrinsic magnetic properties [19-26]. For example, CrN monolayer is predicted to be a planar one atomic-thick structure with robust ferromagnetism [19], or an almost flat hexagonal sheet with minor buckling due to the reduction of a surface dipole moment [20, 21]. Similarly, hexagonal monolayers of chromium pnictides CrX (X=P, As, Sb, Bi) are found to be one atomic-thick flat structure (zero buckling between Cr and X atoms) and ferromagnetically ordered systems with the Curie temperature well above 300 K [22]. At the same time, two-atomic-thick structural CrX (X=P, As) monolayers are also theoretically demonstrated [23]. As the most well-known case of 2D Cr-based materials, $CrI_3$ monolayer is a three-atomic-thick structure, analogue to TMDs, and the ferromagnetic (FM) order can be observed at temperatures below 45 K when few layers are stacking [26]. This finding has provoked a number of experimental and theoretical investigations to try to control the magnetism and realize the field-effect devices based on $CrI_3$ multilayers [27-34]. Both the diversify of structural topology and the eccentric properties revealed by these works demonstrate that 2D Cr-based materials is an interesting platform, even surpass to 2D TMDs, to study the fundamental physics of 2D materials, and has exciting application foreground in various fields. Although some progress has been made in 2D Cr-based materials, the related study and knowledge is still in its infant stage.

Herein we report, for the first time, a new family 2D Cr-based materials --- three-atomic-thick CrY (Y= P, As, Sb) monolayers and their Janus counterparts $Cr_2XY$ (X = P, As, Sb) and $Cr_2NY$. We find that breaking the symmetry (i.e., using N atoms to substitute Y atoms in one side) of pristine CrY monolayers would induce a large out-of-plane intrinsic dipole owing to the electronegativity of N atom much larger than Y atoms, and its related multifunctionalities. Interestingly, $Cr_2NSb$ monolayer is found to be a spin-gapless semiconductor, and functionalizing $Cr_2NP$ can increase the dipole moment and greatly lower work function to 2.46 eV. When $Cr_2NP$ or $Cr_2NHPF$ monolayer is sandwiched between two graphene sheets, interfacial interaction will induce carrier doping of two graphene sheets with different concentrations, thus forming intrinsic *p*-*n* vertical junction. Moreover, Janus $Cr_2NY$ monolayers exhibit excellent electrocatalytic hydrogen evolution activity at N atom sites, even superior to benchmark Pt. The amazing properties making 2D Cr-based materials having exciting applications in nanoelectronics and clean energy conversion.

## 2. Computational method

All the calculations were performed using the first-principles theory within the framework of DFT as implemented in the Vienna Ab Initio Simulation Package (VASP) [35, 36]. The projector augmented wave (PAW) potentials were utilized to deal with the interaction between core electrons [37]. Generalized gradient approximation (GGA) given by Perdew-Burke-Ernzerhof (PBE) was adopted for exchange-correlation functionals [38]. The spin-dependent GGA plus Hubbard U (GGA+U) to approximately described the strongly correlated interactions of the transition metal Cr, where the Hubbard U parameter of the Cr atom was set to 3.0 eV [39, 40]. The screened hybrid Heyd–Scuseria–Ernzerhof (HSE06) functional without Hubbard U correction was used for the calculations of more reliable band gaps [41, 42]. The plane wave cutoff energy was set to be 550 eV， and atomic coordinates were fully relaxed to ensure the total energy in $10^{-5}$ eV precision and force in $10^{-2}$ eV $Å^{-1}$ convergence. The k-points with a 7 × 7 × 1 grid for structure relaxation, and a 9 × 9 × 1 grid for the self-consistent field (SCF) calculations, were sampled by the Gamma-centered Monkhorst–Pack grids. In order to eliminate the coupling force between the periodic images, a vacuum space of 20 Å was inserted along the z direction for the monolayers, and 30 Å for the heterostructures. The van der Waals (vdW) interaction between the monolayers and graphene or H atom were described by the DFT-D3 method [43]. The phonon calculations were carried out by using DFT perturbation theory as implemented in the PHONOPY code [44-46]. A 3 × 3 × 1 supercell was adopted to construct the dynamic matrix and calculate the phonon dispersion. Note that the dipole correction has an ignorable impact on the electronic structures and application properties of the monolayers, thus we did not consider the dipole correction in the following calculations except the calculation of dipole moment and electrostatic potential.

## 3. Results and discussion

### 3.1 The geometric structures and electronic properties

Symmetry is essential to science. In order to explore the effect of symmetry-breaking, we firstly study symmetrical CrY (Y = P, As, Sb) monolayers that all belong to space group *P-nmm* with 2D networks of rectangular sublattices in the x–y plane. Fig.1(a) displays top and side views of CrY monolayers, where the Cr atoms are sandwiched between Y atoms on both sides showing symmetrical structures. In view of exotic properties of Janus monolayers because of breaking structural symmetry, Y atoms on one side of CrY monolayers are replaced with X atoms obtaining Janus $Cr_2XY$ (X, Y = P, As, Sb), as shown in Fig. 1(b), wherein the atomic radius of X atoms are smaller than that of Y atoms. Similarly, $Cr_2NY$ (Y = P, As, Sb) monolayers shown in Fig. 1(c) are obtained by substituting N atoms. Table I lists the optimized lattice constants and bond lengths. For CrY monolayers, the Y-Cr bond length decreases with an increase in the electronegativity of Y atom. Accordingly, the optimized lattice constant of CrY with longer Y-Cr bond is bigger, which are 4.173, 4.262, 4.396 Å for CrP, CrAs and CrSb, respectively. For $Cr_2XY$ cases, the P-Cr, As-Cr, Sb-Cr bond lengths are almost equal to those in CrY monolayers, and the optimized lattice constants is between CrX and CrY ones. But for $Cr_2NY$ cases, due to much greater electronegativity of N atom, the N-Cr bonds are much shorter than P-Cr bond, thus the optimized lattice constants of $Cr_2NY$ monolayer are smaller than all of CrY ones.

The formation energy ($E_f$) calculations are performed to investigate the energetic stability as:

$$E_f = E_{total} - \sum_{i=1}^{4} E_i$$

where $E_{total}$ represents the total cell energy; $E_i$ refers to the chemical potential of each atom of the unit cell, which is the normalized energy of the corresponding bulk phase. It can be seen from Table SI (see the supplemental material [47]) that the formation energy of the monolayers are negative values showing energetic stability. To determine dynamical stability, the phonon dispersion is calculated and shown in Fig. S1 (see the supplemental material [47]), which includes 3 acoustic and 9 optical branches corresponding to four atoms in each unit cell for every structure, and the phonon modes are free from any imaginary frequencies throughout the whole Brillouin zone, indicating the intrinsic structural stability of CrY, $Cr_2XY$ and $Cr_2NY$ monolayers. However, CrN monolayer is found to be unable (Fig. S2(a)). It is also necessary to assess the effect of lattice distortion on structural stability. To guarantee the positive-definiteness of strain energy following lattice distortion, the linear elastic constants of a stable crystal have to obey the Born–Huang criteria. For a mechanically stable 2D monolayer with tetragonal symmetry, the elastic constants need to satisfy $C_{11} > |C_{12}|$ and $C_{66} > 0$. We calculated the elastic constants of CrY, $Cr_2XY$ and $Cr_2NY$ monolayers, and found that except $Cr_2AsSb$ (the $Cr_2AsSb$ monolayer is not focused below), others can satisfy the conditions (Table SI), indicating the monolayers except $Cr_2AsSb$ are mechanically stable.

Structural symmetry-breaking can significantly influence physical properties of 2D materials [4, 5]. To testify this, dipole moments of all monolayers are calculated and listed in Table I. In Janus $Cr_2XY$ monolayers, Cr atoms are coordinated with X and Y atoms forming trigonal prisms with mirror asymmetry, leading to a dipole pointing from Y to X atoms, and dipole moments are 0.05, 0.12, and 0.09 D for $Cr_2PAs$, $Cr_2PSb$ and $Cr_2AsSb$, respectively (the dipole disappears in symmetrical CrY monolayers). Things are more interesting in $Cr_2NY$ monolayers that the greater electronegativity difference between N and Y atoms results in larger dipole moments, which are 0.39, 0.33 and 0.26 D for $Cr_2NP$, $Cr_2NAs$ and $Cr_2NSb$, respectively. Therefore, breaking symmetry by substituting N atoms can effectively increase the dipole moment, especially for $Cr_2NP$ which is more than two times in that of MoSSe (0.18 D) [48, 49].

Generally, the dipole of 2D Janus monolayers, such as $Cr_2XY$ and MoSSe, points from one side of atom with lower electronegativity to another side. Unexpectedly, opposite situation is found in $Cr_2NY$ monolayers, where the dipole pointing from N to Y atoms where Y atom possess lower electronegativity than N atom. This anomalous behavior in $Cr_2NY$ can be proved as follows. The net vertical electric field generated from the dipole can induce a difference in electrostatic potential on both sides of Janus monolayers, and the side with higher potential is always the end where the dipole points to. Thus, to further verify the size and direction of the dipole, we can investigate electrostatic potentials in the direction perpendicular to the monolayers, as shown in Fig. 2. For CrY monolayers, it aligns to the vacuum energy in both ends because there is no dipole in these symmetrical structures. For $Cr_2XY$ monolayers, the potentials are 0.12, 0.23 and 0.17 eV lower in Y sides of $Cr_2PAs$, $Cr_2PSb$ and $Cr_2AsSb$, respectively, where Y atoms possess lower electronegativity. Whereas, in each $Cr_2NY$ monolayer, the side of N atoms which possesses higher electronegativity have lower potential than another side, where potential differences are 1.18, 0.77 and 0.78 eV for $Cr_2NP$, $Cr_2NAs$ and $Cr_2NAs$, respectively.

To explain the phenomenon above, we propose a simplified model analyzing local dipole. The intrinsic dipoles are calculated by dipole correction, but the method can't be used to deal with the local dipole. A solution is to divide the three-atom-layer structure into two parts which are from Cr to atoms on both sides (Fig 3), and then the local dipoles can be roughly solved by Bader charges and atomic layer spacing between Cr and one of both sides, which can be used to figure out the root of the larger dipole with abnormal direction for $Cr_2NY$. Here, the two local dipoles can be written as:

$$\mu_i = \Delta q_i \cdot \Delta z_i;\ (i = 1, 2)$$

where $\mu_i$, $\Delta q_i$ and $\Delta z_i$ is local dipole, Bader charges and atomic layer spacing along z direction between Cr and one of both sides, which are calculated and summarized in Table SII (see the supplemental material [47]). For CrY cases, $\mu_1$ and $\mu_2$ have equal values but opposite direction, because identical Y atoms lead to equal charge transfer ($\Delta q_1$ and $\Delta q_2$) and equal spacing ($\Delta z_1$ and $\Delta z_2$) between Cr and Y on both sides. Fig. 3(a) shows the result of CrP monolayer indicating there is no dipole, and the situation is the same for CrAs and CrSb cases. On the contrary, for $Cr_2XY$ and $Cr_2NY$ monolayers, charge transfer and atomic layer spacing from Cr to two sides are different due to their different electronegativity, leading to different $\mu_1$ and $\mu_2$, where $\mu_1$ ($\mu_2$) can be defined as pointing to atoms with higher (lower) electronegativity. For $Cr_2XY$ monolayers, the local dipole $\mu_1$ has bigger value than $\mu_2$ because of the larger $\Delta q_1$, leading to the direction of dipole from Y to X atoms. Note that the actual intrinsic dipole, $\mu$, in $Cr_2XY$ monolayers is small owing to the small differences of charge transfer and atomic layer spacing on both sides; $Cr_2PAs$ monolayer is take as an example as shown in Fig. 3(b). For $Cr_2NY$ monolayers, however, the atomic layer spacing $\Delta z_1$ (Cr-N) is much shorter than $\Delta z_2$ (Cr-Y), thus the local dipole $\mu_1$ is much smaller than $\mu_2$. As a result, the intrinsic dipole, $\mu$, in $Cr_2NY$ monolayers is large and points from N to Y atoms. For example, focusing on $Cr_2NP$ (Fig. 3(c)), although the charge transfer $\Delta q_1$ is nearly twice as large as $\Delta q_2$, the atomic layer spacing $\Delta z_1$ is about a third shorter than $\Delta z_2$, resulting in the local dipole $\mu_1$ much smaller than $\mu_2$. Therefore, $Cr_2NY$ monolayers have large dipole with opposite direction which rely on electronegativity differences between N and Y atoms.

To determine the magnetic ground state, we carry out calculations of the energy differences ($E_{AFM\text{-}FM}$) between ferromagnetic (FM) and antiferromagnetic (AFM) states, where AFM state is briefly set the opposite spin direction for Cr atoms in unit cell. The positive values of $E_{AFM\text{-}FM}$ (Table I) pronounce the FM ground states for all CrY, $Cr_2XY$ and $Cr_2NY$ monolayers. In the monolayers, each Cr atom gives 3 electrons to form ionic bonding with the ligands and leaves 3 electrons, so there is a large magnetic moment of 6 μB in each unit cell.

As listed in Table I, the bandgap energies calculated by the GGA+U method are less than HSE06 hybrid functional method, because the independent electron picture breaks down as a result of strong Coulombic interactions [50-52]. To investigate electronic properties of the FM ground states of these monolayers, we focus on the spin-polarized electronic band structures calculated by the HSE06 hybrid functional method. According to the characteristics of the structures, we choose the high symmetry $k$-path in the Brillouin zone (Fig. S2(b)). Fig. 4 shows the bands projected onto spin-up and spin-down, respectively represented by red and blue lines. For CrY monolayers (top part in Fig. 4), it is found that the band of CrSb monolayer is metallic, but the CrP and CrAs ones are semiconductors with indirect-gap of 0.543 and 0.437 eV, whose valence-band maximum (VBM) and conduction-band minimum (CBM) edges locate at the Γ and S points of the BZ, respectively. Same as the CrP and CrAs monolayers, $Cr_2XY$ monolayer are also indirect semiconductors with the band gap ranging from 0.140 to 0.607 eV (middle part in Fig. 4). However, because the charges more dramatically transfer from Cr to N atoms after N atomic substitution significantly affecting the electronic structure, the situations are completely different in $Cr_2NY$ monolayers (lower part in Fig. 4). Firstly, $Cr_2NP$ keeps indirect gap with CBM and VBM edges residing at the S and Γ points, while $Cr_2NAs$ become a direct gap semiconductor with the CBM and VBM edges both locating at S point. The energy gaps of $Cr_2NP$ and $Cr_2NAs$ are 0.818 and 0.872 eV, respectively, which are much bigger than those of CrY and $Cr_2XY$. Things are more interesting in $Cr_2NSb$ that the CBM and VBM edges meet at S point leading to a very small energy gap close to 0 eV, and both CBM and VBM come from spin-up bands, indicating that the $Cr_2NSb$ monolayer is a 2D spin-gapless semiconductor (SGS) [53-55]. The electron excitation behavior with only spin-up carriers in SGS always can easily happen, suggesting that the $Cr_2NSb$ monolayer has potential applications in spintronics.

### 3.2 The *p-n* vertical junctions induced by symmetry-breaking

Special structures of $Cr_2NY$ monolayers result in a large dipole, especially for $Cr_2NP$. To explore their applications, we respectively use CrP and $Cr_2NP$ as an intermediate layer to sandwich in-between two layers of graphene (G), forming van der Waals (vdW) heterostructures G/CrP/G (Fig. 5(a)) and G/$Cr_2NP$/G (Fig. 6(a)). Here, in order to minimize the lattice mismatch between CrP ($Cr_2NP$) and graphene layers, we have considered $\sqrt{5}\times\sqrt{5}\times 1$ and $\sqrt{15}\times\sqrt{15}\times 1$ supercell of CrP ($Cr_2NP$) and graphene, respectively. All the atomic positions and lattice vectors of heterostructures are fully relaxed, and we obtain a vertical interlayer distance of 3.43 Å for G/CrP/G, with no chemical bonds at interface region indicating the existence of a weak van der Waals (vdW) interaction. For G/$Cr_2NP$/G heterostructure, the vdW interaction between $Cr_2NP$ and graphene is stronger (weaker) in N (P) atom side because of the shorter (longer) layer distance of 3.35 Å (3.53 Å). The energetic stability is examined by bonding energy $E_b$ comparing the total energies of the final heterostructure with ones of the isolated components:

$$E_b = E_H - E_{in} - E_{G1} - E_{G2}$$

where $E_H$, $E_{in}$, $E_{G1}$ and $E_{G2}$ and are the energies of heterostructure, isolated intermediate layer and graphene on two sides. It is found that the formation of G/CrP/G (G/$Cr_2$NP/G) is an exothermic process with $E_b$ of -2.29 eV (-2.36 eV), indicating their energetic stability.

Fig. 5(b) shows both induced p-type doping of two graphene sheets in G/CrP/G, because the Dirac points of graphene are the same for energy and both higher than the Fermi level from electronic bands and density of states (DOS) projected to graphene layers (Fig. 5(c)). Moreover, interfacial interaction is analyzed by charge density difference, as shown in Fig. 5(d), where the carriers from CrP to two graphene sheets are holes, further proving the same p-type doping of graphene sheets. For the G/$Cr_2$NP/G case, due to the intrinsic dipole of Janus $Cr_2$NP, two graphene sheets are induced different doping types (Fig. 6(b)). In Fig. 6(c), the red (blue) shows the band or DOS projected onto graphene sheet on N (P) side, which displays n-type (p-type) doping for graphene with Dirac point being lower (higher) than the Fermi level. The Dirac point difference of 0.79 eV can be used to describe difference of doping between two graphene sheets. The phenomenon derives from interfacial interaction between graphene and Janus $Cr_2$NP. From the charge density difference as shown in Fig. 6(d), because of the built-in vertical electric field generated from dipole of $Cr_2$NP monolayer, the electrons (holes) can be transferred from intermediate layer to the graphene sheets by the N (P) side forming n-type (p-type) doping of graphene. Therefore, an ultrathin p-n vertical junction is naturally formed [56, 57], indicating that symmetry-breaking is an effective strategy to make 2D materials such exciting and useful in nanoelectronics.

The strategy of functionalization can further break symmetry to impact properties of 2D materials [58, 59]. We next consider that hydrogenating and fluorinating at N ($Cr_2$NHP/$Cr_2$NFP) or P ($Cr_2$NPH/$Cr_2$NPF) atoms to tune the electronic properties, which are stable composite structures except the $Cr_2$NFP monolayer. The Bader charges and dipole moments are calculated and listed in Table SIII. It is found that the electrons transfer from P to F atoms in $Cr_2$NPF monolayer, because the electronegativity of F atom is greater than that of P atom; and the electrons transfer from P to H atoms in $Cr_2$NPH and from H to N atoms in $Cr_2$NHP monolayer, due to the electronegativity of H atom greater than P atom but smaller than N atom. Interestingly, the dipole moments are enhanced in $Cr_2$NHP and $Cr_2$NPF monolayers, but decreased and even converted to opposite direction for $Cr_2NPH$ monolayer. Affected by the changes of dipole, electrostatic potential differences shown in Fig. 7 become larger in $Cr_2NPF$ and $Cr_2NHP$, but become negative in $Cr_2NPH$, which is consistent with the previous conclusion in situations of Janus $Cr_2XY$ and $Cr_2NY$. Moreover, an interesting result of $Cr_2NHP$ monolayer is that hydrogenating can lower its work function to 2.46 eV. In surface of photoelectric devices, capture and excitation of charge are extremely necessary and a lower work function can make them easily achieved, indicating that $Cr_2NHP$ monolayer is a potential electron emission layer material.

We also consider that functionalization on both sides of $Cr_2NP$ monolayer, including two cases of $Cr_2NHPH$ and $Cr_2NHPF$, which can be viewed as functionalization at P atom in $Cr_2NHP$. As listed in Table SIII, the dipole moment is weakened when hydrogenating, and enhanced when fluorinating compared with $Cr_2NHP$, and the electrostatic potential differences (Fig. 7) increase or decrease accordingly. Thus, a large dipole moment up to 1.53 D is obtained in $Cr_2NHPF$ monolayer. A $\sqrt{5} \times \sqrt{5} \times 1$ supercell of $Cr_2NHPF$ is sandwiched in two graphene sheets forming vdW heterostructures G/$Cr_2NHPF$/G (Fig. 8(a)), where layer distances are shorter on both sides compared with G/$Cr_2NP$/G, indicating stronger interfacial interaction. The $E_b$ of -2.36 eV can further prove its energetic stability. Like G/$Cr_2NF$/G case, induced doping principle of G/$Cr_2NHPF$/G depicted in Fig. 8(b) is the same. As expected, as shown in Fig. 8(c), more obvious doping can be found in graphene sheets of G/$Cr_2NHPF$/G with the Dirac point difference increasing to 1.02 eV because of the larger vertical electric field. The stronger interfacial interaction can be confirmed by the charge density difference (Fig. 8(d)), where the carrier transfers from $Cr_2NHPF$ to graphene on fluorinated P atom (hydrogenated N atom) side is net electronic (hole). Therefore, based on G/$Cr_2NHPF$/G, an ultrathin p-n vertical junction with larger carrier concentration can be potential prepared, which has great potential in nanoelectronics.

### 3.3 Hydrogen evolution reaction performance induced by symmetry-breaking

We finally investigate symmetry-breaking-induced hydrogen evolution reaction (HER) performance of 2D Cr-based materials. The electrocatalytic activities for HER are investigated in $2 \times 2 \times 1$ supercell of the monolayers by calculating Gibbs free energy difference ($\Delta G$). The Sabatier principle states that the interaction between the catalyst and H atoms should be neither too strong nor too weak [60, 61]. Here, we find that the ΔG has large positive values (Fig. S3) of P, As and Sb atom sites in CrY and $Cr_2XY$ monolayers, indicating that the interaction between the monolayers and H atom is weak, and the desorption reaction can more easily happen but adsorption reaction would be restricted. On the contrary, the N atom sites in $Cr_2NY$ monolayers display excellent HER activities. As shown in Fig. 9(a), the ΔG has small values of 0.155, 0.049 and 0.024 eV of N atom sites in $Cr_2NP$, $Cr_2NAs$ and $Cr_2NSb$ monolayers, respectively; indicating their good catalytic activity, even better than Pt (|ΔG| = 0.09 eV) for $Cr_2NAs$ and $Cr_2NSb$ cases. Interestingly, we note that the HER catalytic activity mainly depends on electronegativity of adsorption sites: ΔG has smaller values in sites with higher electronegativity. Thereby, P, As, and Sb atom sites are catalytically inactive, while N atom sites have better activities. On the other hand, it is not negligible for the atoms of non-adsorption sites: the adsorption sites have better activities when the non-adsorption sites possess lower electronegativity. Therefore, in $Cr_2NY$ monolayers, the N atom sites in $Cr_2NSb$ monolayer have the best HER performance due to the Sb atom has the lowest electronegativity.

To explore the mechanism of HER activity difference in 2D Cr-based materials, Bader charges of adsorbing sites in slab monolayers are calculated and summarized in Fig. S3. It is found that the adsorbing sites with higher electronegativity would obtain more electrons, and its amount decreases (increases) with an increase (decrease) electronegativity of non-adsorption sites. The atom sites accumulating more charges have stronger interactions with H atom, which means that more stable adsorbed states can be formed and the difficulty of desorption reaction can be reduced. Thus, the N atom sites obtaining more electrons have the best HER activity, and the $Cr_2NSb$ monolayer is optimal HER catalyst because Sb atom with the lowest electronegativity can promote the accumulation of charges in N atom site.

It has been demonstrated that external strain is an effective strategy to tune the HER performance of catalysts [62-66]. We apply biaxial tensile strains along the plane of the CrY and $Cr_2XY$ monolayers, and find that P atom sites become HER active. Fig. 9(b) shows the ΔG change of P atom sites as a function of the applied biaxial strain, where the case of Pt is taken as catalytic window highlighted in green. One can see that when the tensile strain is from 3% to 5%, the HER activity of P atom sites can be tuned to the catalytic window, thereby possessing better HER activities than that of Pt. Moreover, the values of ΔG close to 0 at the tensile strain of about 4%, leading to optimal HER performances of P atom sites in the CrY and $Cr_2XY$ monolayers. Therefore, by breaking the structural symmetry via N atom substitution, the $Cr_2NY$ monolayers have excellent HER activity, and applying biaxial tensile strain, 2D Cr-based materials become HER active.

## 4. Conclusions

In summary, we propose a series of Janus $Cr_2NY$ (Y=P, As, Sb) monolayers with multiple functionalities by breaking structural symmetry of CrY. The results show that the geometric and electronic structures can be significantly regulated, which result in larger dipole and further make great potentials in various applications for $Cr_2NY$. Specifically, $Cr_2NSb$ possesses fascinating electronic band which is a spin-gapless semiconductor. Through the strategy of functionalization further breaking symmetry of $Cr_2NP$, work function can be lowered to 2.46 eV and dipole moment can be enhanced to 1.53 D. Sandwiching either $Cr_2NP$ or functionalized $Cr_2NP$ between two graphene sheets, interfacial interaction will induce different carrier doping types of graphene sheets generating an ultrathin p-n junction. Moreover, structural symmetry-breaking can promote the accumulation of charges in N sites of $Cr_2NY$, leading to optimal hydrogen evolution performance especially for $Cr_2NSb$. Through breaking symmetry, we rationally design multifunctional 2D Janus materials with exciting applications in nanoelectronics and clean energy conversion.

## Acknowledge

This work was supported by the National Natural Science Foundation of China (Grants No. 52172088 and No. 51772085) and Natural Science Foundation of Hunan Province (No. 2020JJ4190 and No. 2021JJ30112).

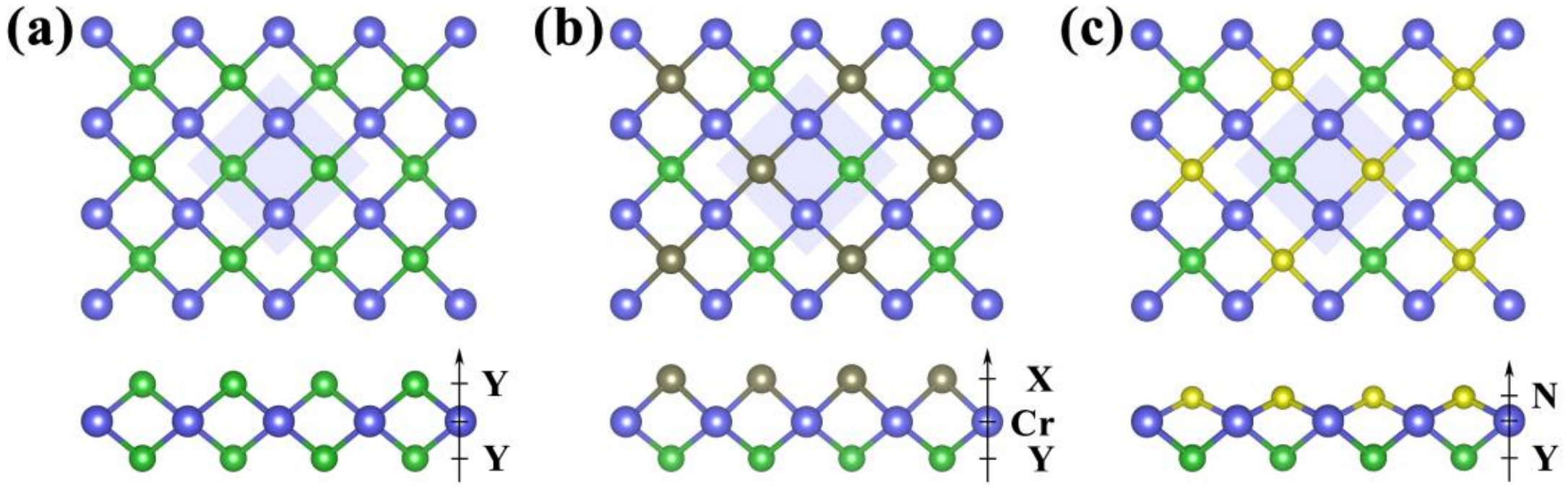

FIG. 1. Top (top panel) and side (lower panel) views of (a) CrY, (b) $Cr_2XY$, (c) $Cr_2NY$ monolayers (X, Y = P, As, Sb). The blue regions represent the unit cells; the yellow, green, brown and blue balls represent N, Y, X, and Cr, respectively.

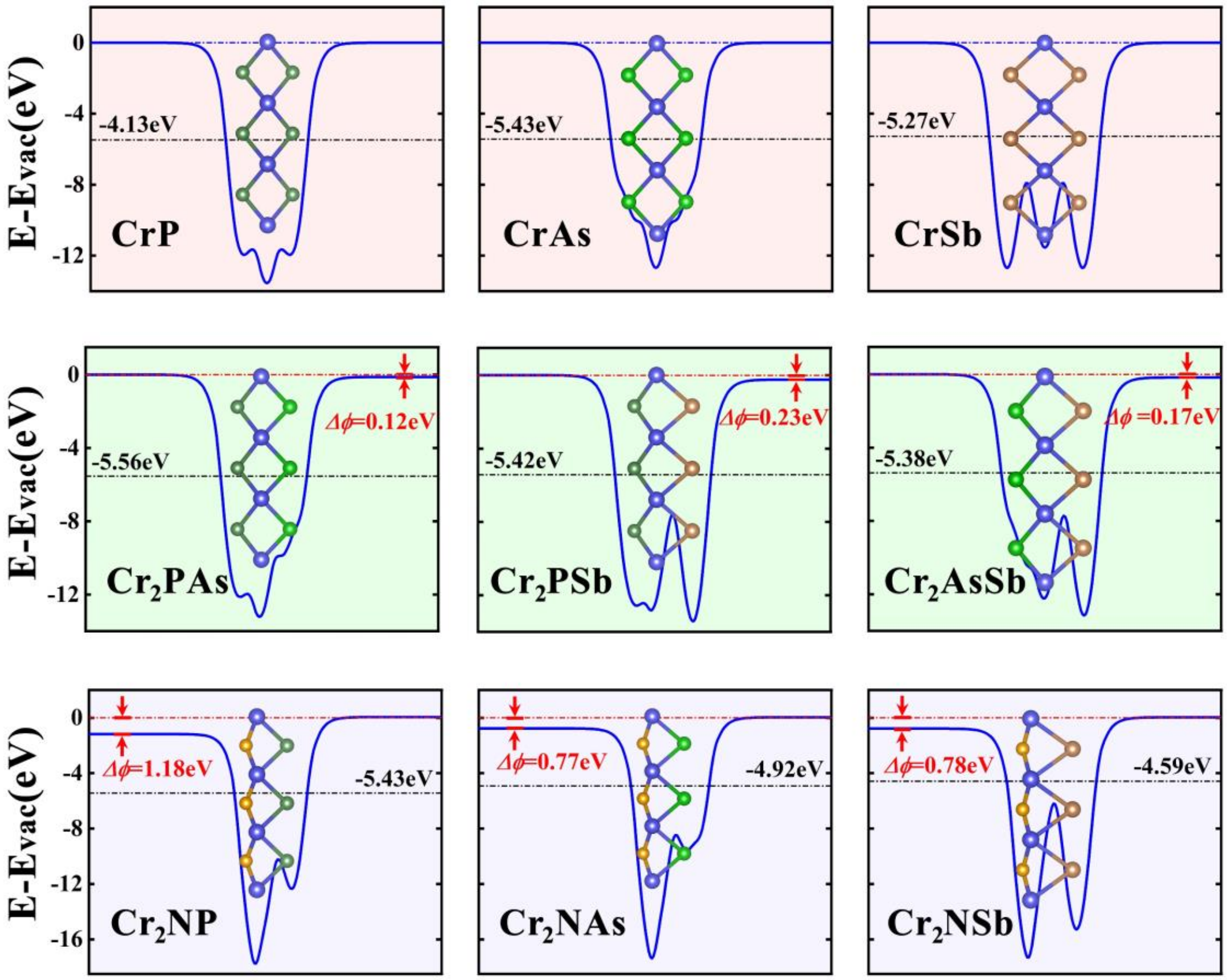

FIG. 2. Electrostatic potential with respect to the vacuum energy in the direction perpendicular to CrY, $Cr_2XY$ and $Cr_2NY$ monolayers. The red and black energy values are electrostatic potential differences and Fermi level; the yellow, dark green, green, brown and blue balls in the structures represent N, P, As, Sb and Cr, respectively.

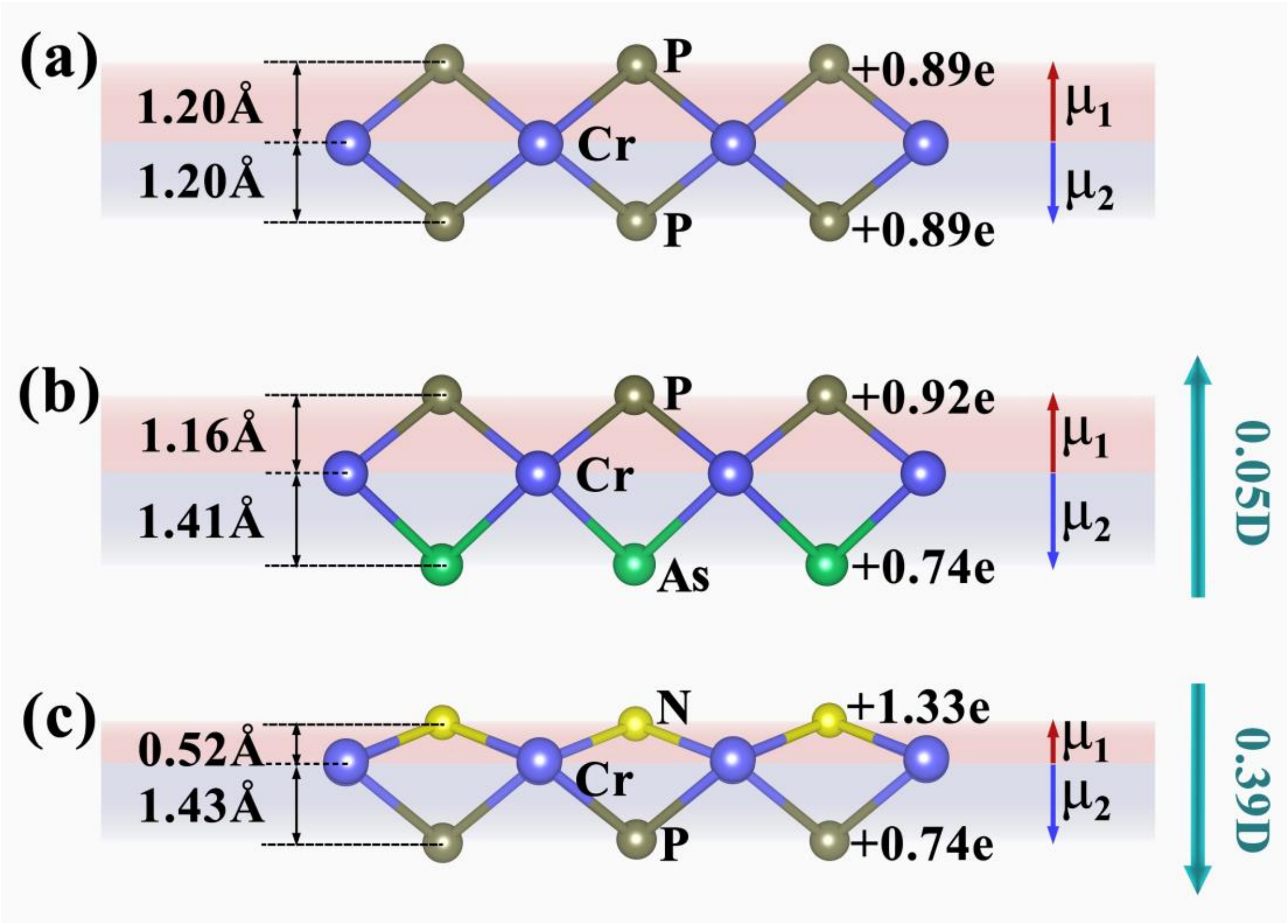

FIG. 3. Dipole schematic diagram of (a) CrP, (b) $Cr_2PAs$ and (c) $Cr_2NP$ monolayers. The pink and purple regions show local dipole with different directions; the black numbers show transferred charge and spacing along z direction on both sides; the green arrows and numbers show actual dipole moment.

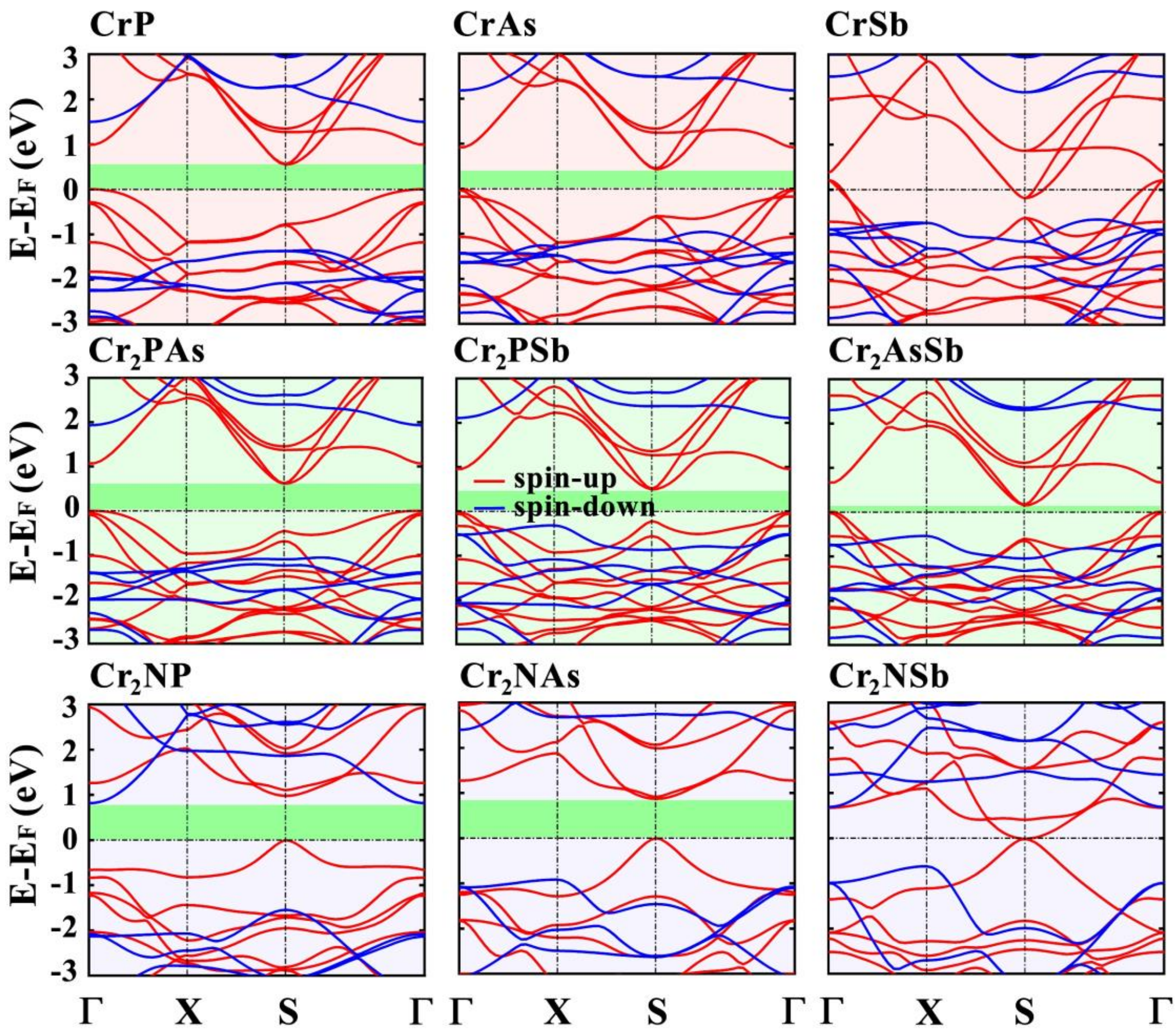

FIG. 4. The spin-polarized electronic band structures calculated by the HSE functional with respect to the $E_F$ of the CrY, $Cr_2XY$ and $Cr_2NY$ monolayers. The red and blue lines represent bands projected to spin-up and spin-down; the energy gap is highlighted in green.

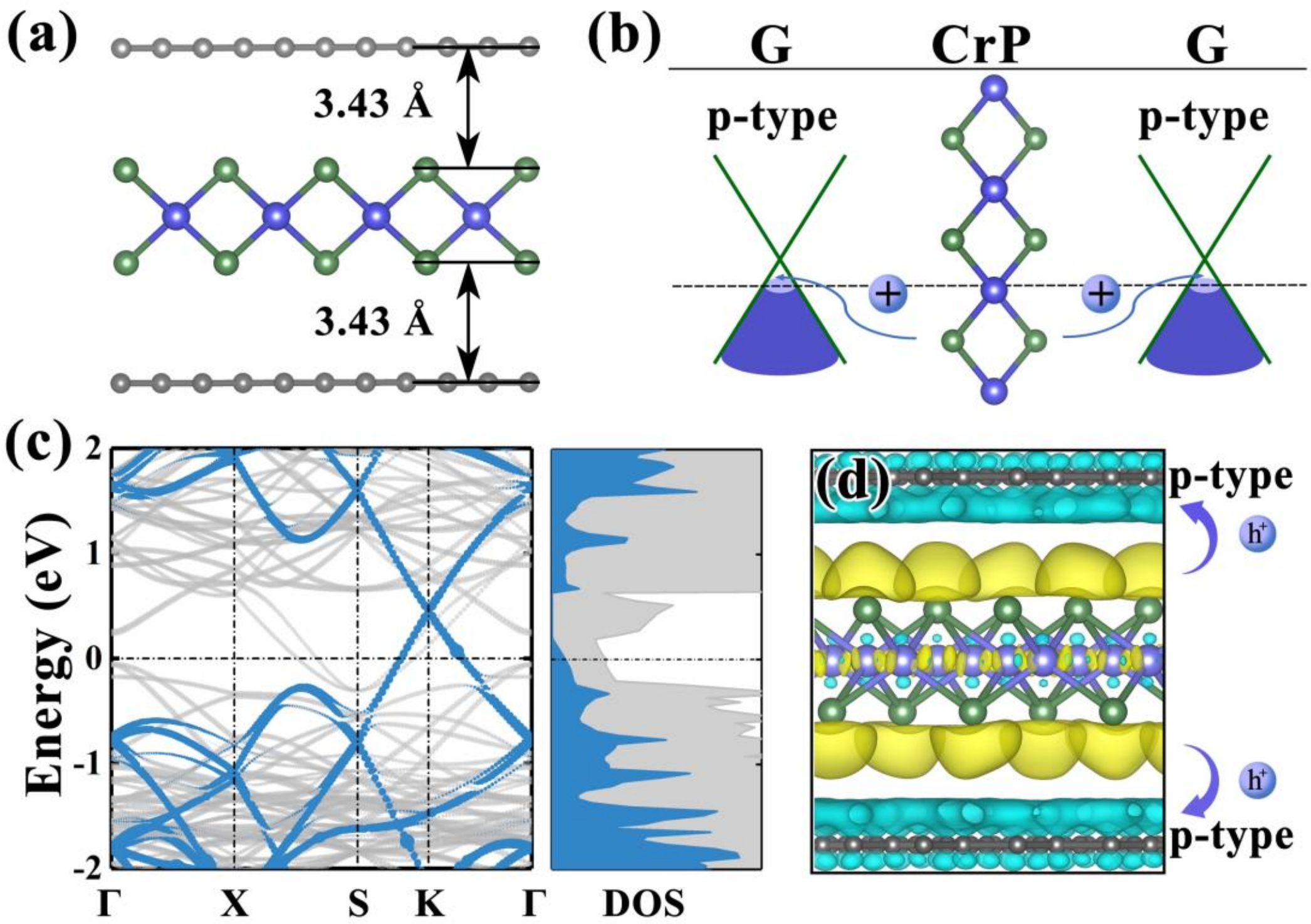

FIG. 5. (a) Structure of vdW heterostructure G/CrP/G. The black numbers show the vertical interlayer distances. (b) Induced doping schematic diagram of G/CrP/G. (c) Electronic band structure and DOS. The blue shows the band projected onto two layers of graphene, which are completely overlapping; the gray shows the band projected onto CrP in band structure and total density of state in DOS. (d) The charge density difference with an isovalue of 0.0003 e/Å$^3$. The yellow and blue regions indicate the accumulation and depletion of electrons, respectively.

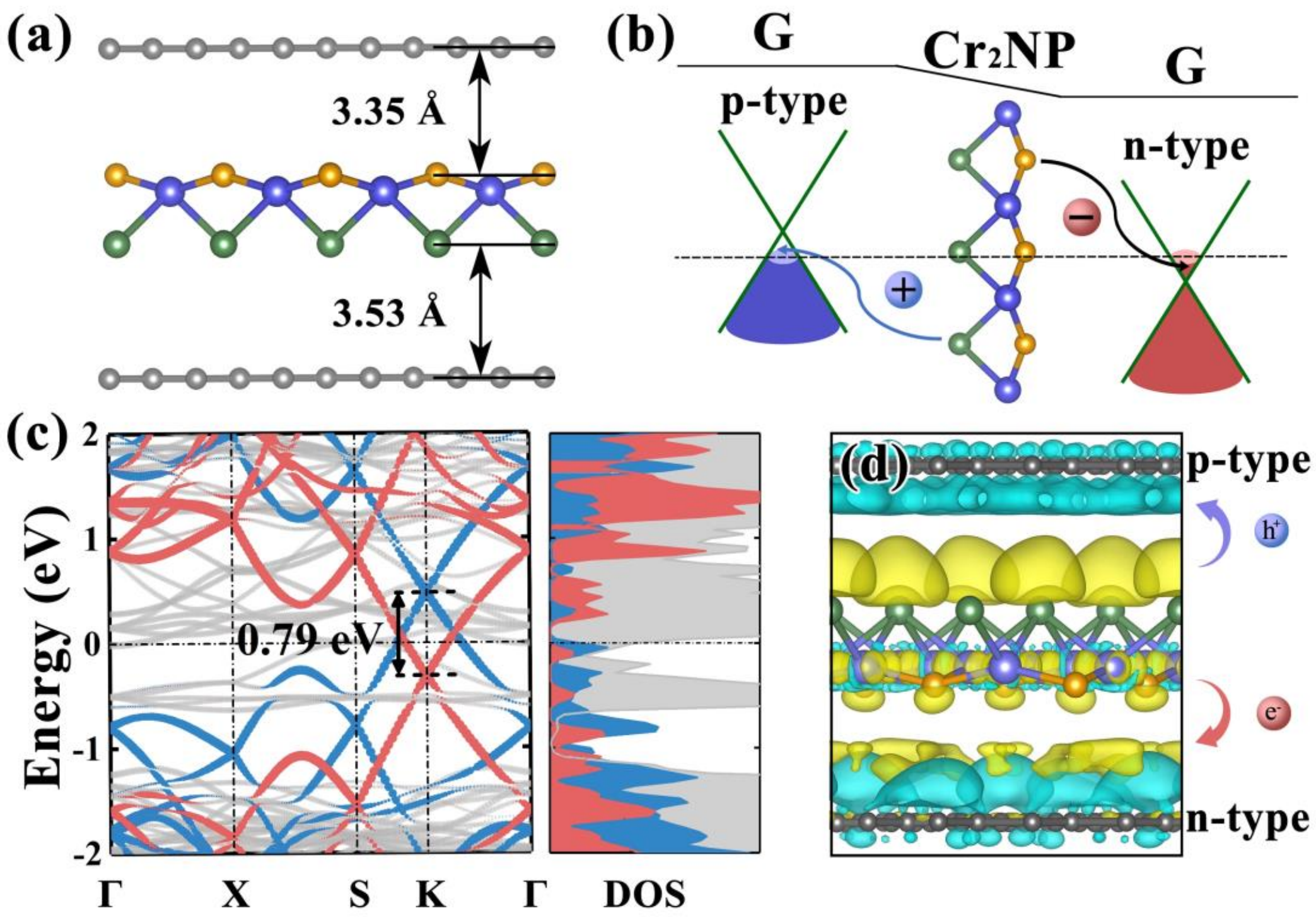

FIG. 6. (a) Structure of vdW heterostructure G/$Cr_2NP$/G. The black numbers show the vertical interlayer distances. (b) Induced doping schematic diagram of G/$Cr_2NP$/G. (c) Electronic band structure and DOS. The red (blue) shows the band projected onto the graphene layer on side of N (P) atoms; the gray shows the band projected onto $Cr_2NP$ in band structure and total density of state in DOS. (d) The charge density difference with an isovalue of 0.0003 e/Å$^3$. The yellow and blue regions indicate the accumulation and depletion of electrons, respectively.

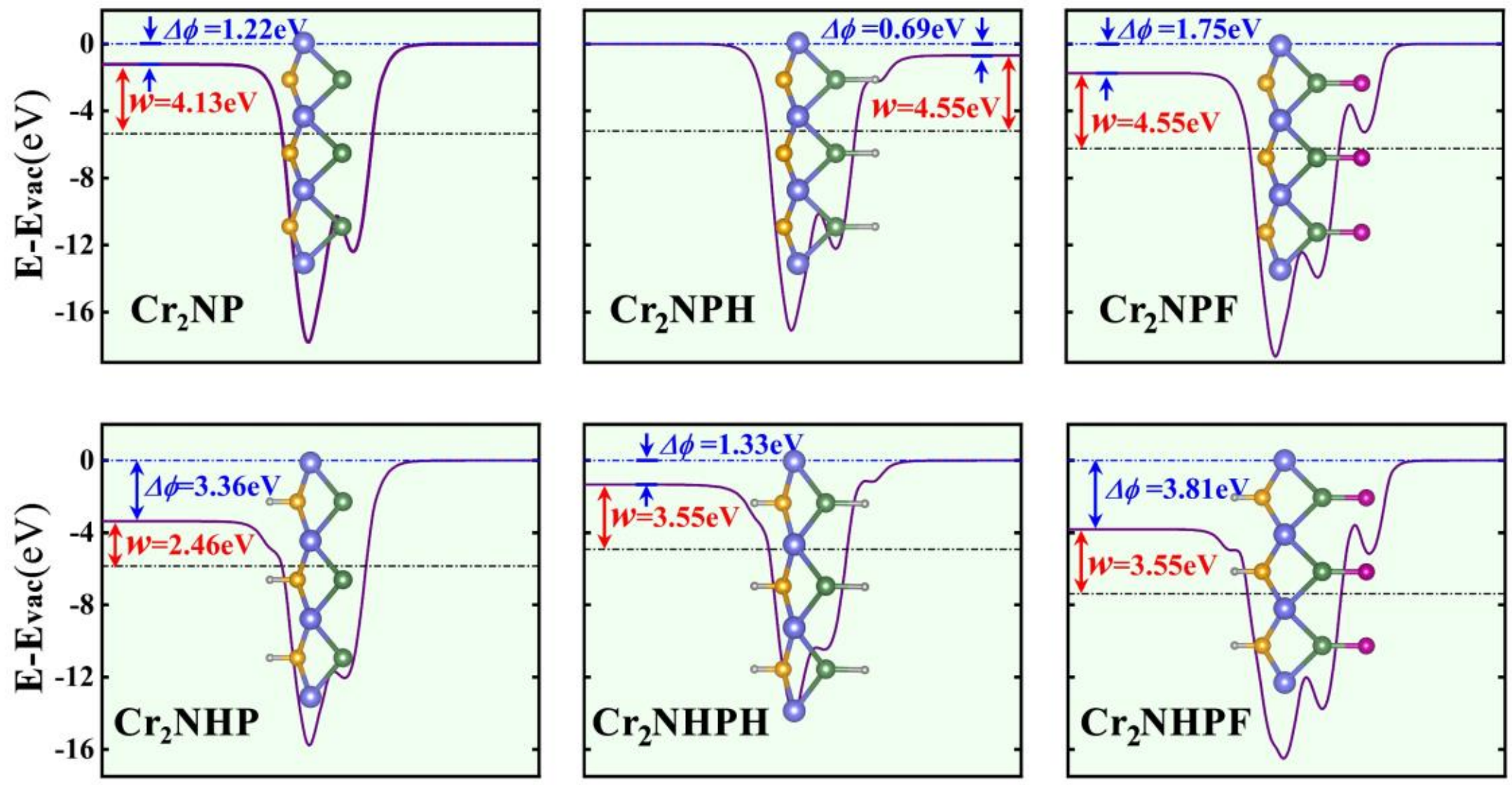

FIG. 7. Electrostatic potential of functionalized $Cr_2NP$ in the direction perpendicular to the layers. Electrostatic potential difference (*Δϕ*) between two sides is indicated by blue arrow; the work function (*w*) of the side with lower potential is shown by red arrow.

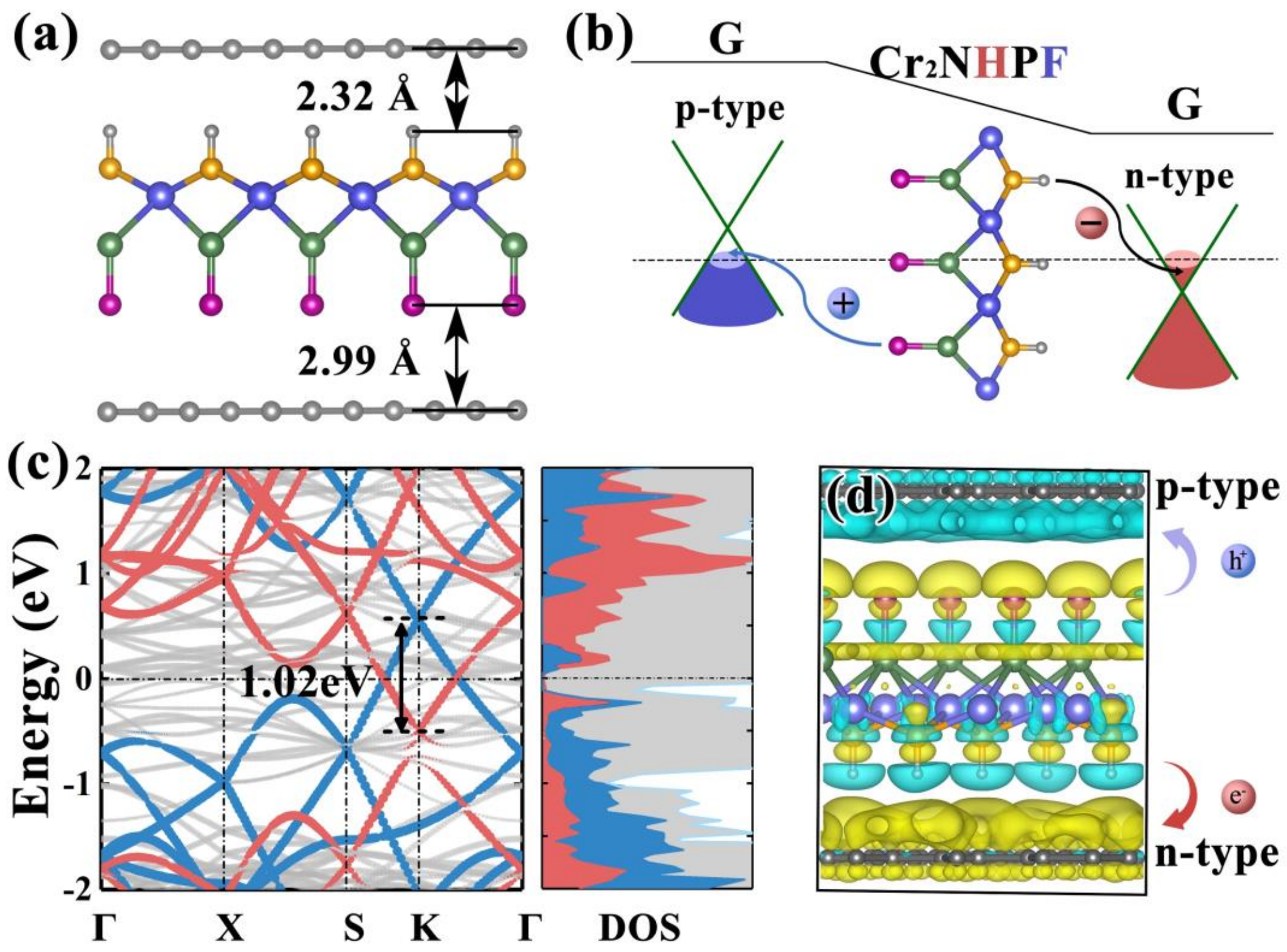

FIG. 8. (a) Structure of vdW heterostructure G/ $Cr_2NHPF$ /G. The black numbers show the vertical interlayer distances. (b) Induced doping schematic diagram of G/$Cr_2NHPF$/G. (c) Electronic band structure and DOS. The red (blue) shows the band projected onto the graphene layer on side of H (F) atoms; the gray shows the band projected onto $Cr_2NP$ in band structure and total density of state in DOS. (d) The charge density difference with an isovalue of 0.0003 e/Å$^3$. The yellow and blue regions indicate the accumulation and depletion of electrons, respectively.

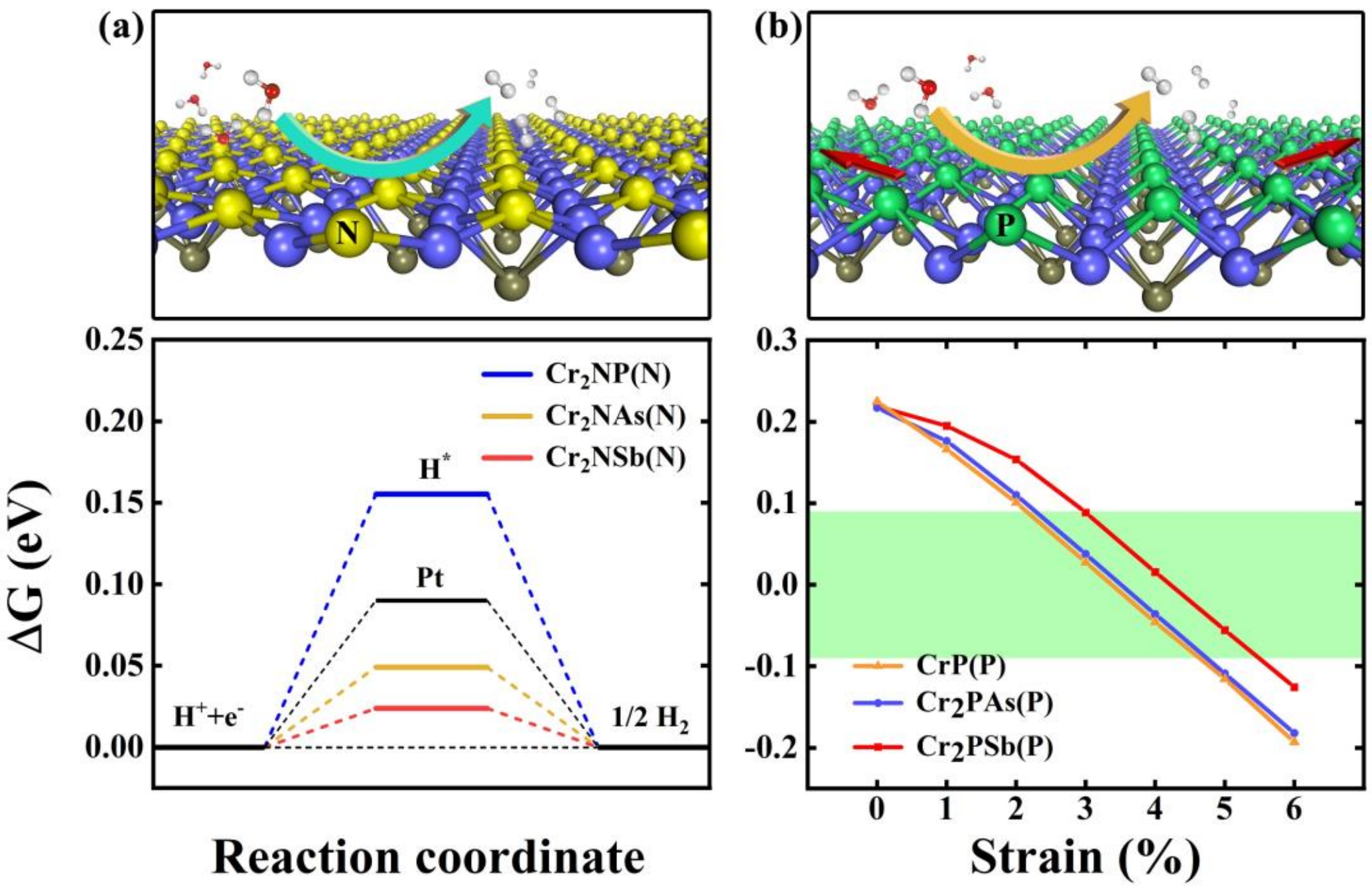

FIG. 9. Schematic diagram (top panel) and the Gibbs free energy differences (lower panel) of a hydrogen atom adsorbed at (a) N sites of $Cr_2NY$ monolayers and (b) P sites of structures applied biaxial strain.

TABLE I. Optimized lattice constant ($a$), bond length ($d$), dipole moment ($\mu$), exchange energies ($E_{AFM\text{-}FM}$), band gaps by GGA+U ($E_g^{GGA+U}$) and HSE ($E_g^{HSE}$) of CrX, $Cr_2XY$ and $Cr_2NY$ monolayers.

| Monolayer | $a$(Å) | $d_{X\text{-}Cr}$(Å) | $d_{Y\text{-}Cr}$(Å) | $\mu$(D) | $E_{AFM\text{-}FM}$ (meV) | $E_g^{GGA+U}$ (eV) | $E_g^{HSE}$ (eV) |
|---|---|---|---|---|---|---|---|
| CrP | 4.173 | - | 2.409 | 0 | 700.8 | 0.041 | 0.543 |
| CrAs | 4.262 | - | 2.528 | 0 | 676.6 | 0.067 | 0.437 |
| CrSb | 4.396 | - | 2.752 | 0 | 497.7 | 0 | 0 |
| $Cr_2PAs$ | 4.216 | 2.406 | 2.532 | 0.05 | 676.1 | 0.145 | 0.607 |
| $Cr_2PSb$ | 4.289 | 2.400 | 2.757 | 0.12 | 480.2 | 0.126 | 0.514 |
| $Cr_2AsSb$ | 4.329 | 2.522 | 2.757 | 0.09 | 522.6 | 0 | 0.140 |
| $Cr_2NP$ | 3.873 | 2.005 | 2.406 | 0.39 | 839.9 | 0.081 | 0.818 |
| $Cr_2NAs$ | 3.913 | 2.010 | 2.529 | 0.33 | 427.4 | 0 | 0.872 |
| $Cr_2NSb$ | 3.942 | 2.005 | 2.768 | 0.26 | 62.08 | 0 | 0.001 |

# Supplemental Material

## Symmetry-breaking-induced multifunctionalities of two-dimensional chromium-based materials for nanoelectronics and clean energy conversion

Lei Li[1], Tao Huang[1], Kun Liang[1], Yuan Si[1], Ji-Chun Lian[1], Wei-Qing Huang[1]*, Wang-Yu Hu[2], Gui-Fang Huang[1#]

1 Department of Applied Physics, School of Physics and Electronics, Hunan University, Changsha 410082, China

2 School of Materials Science and Engineering, Hunan University, Changsha 410082, China

## AUTHOR INFORMATION

*. Corresponding author. *E-mail address:* wqhuang@hnu.edu.cn

#. Corresponding author. *E-mail address:* gfhuang@hnu.edu.cn

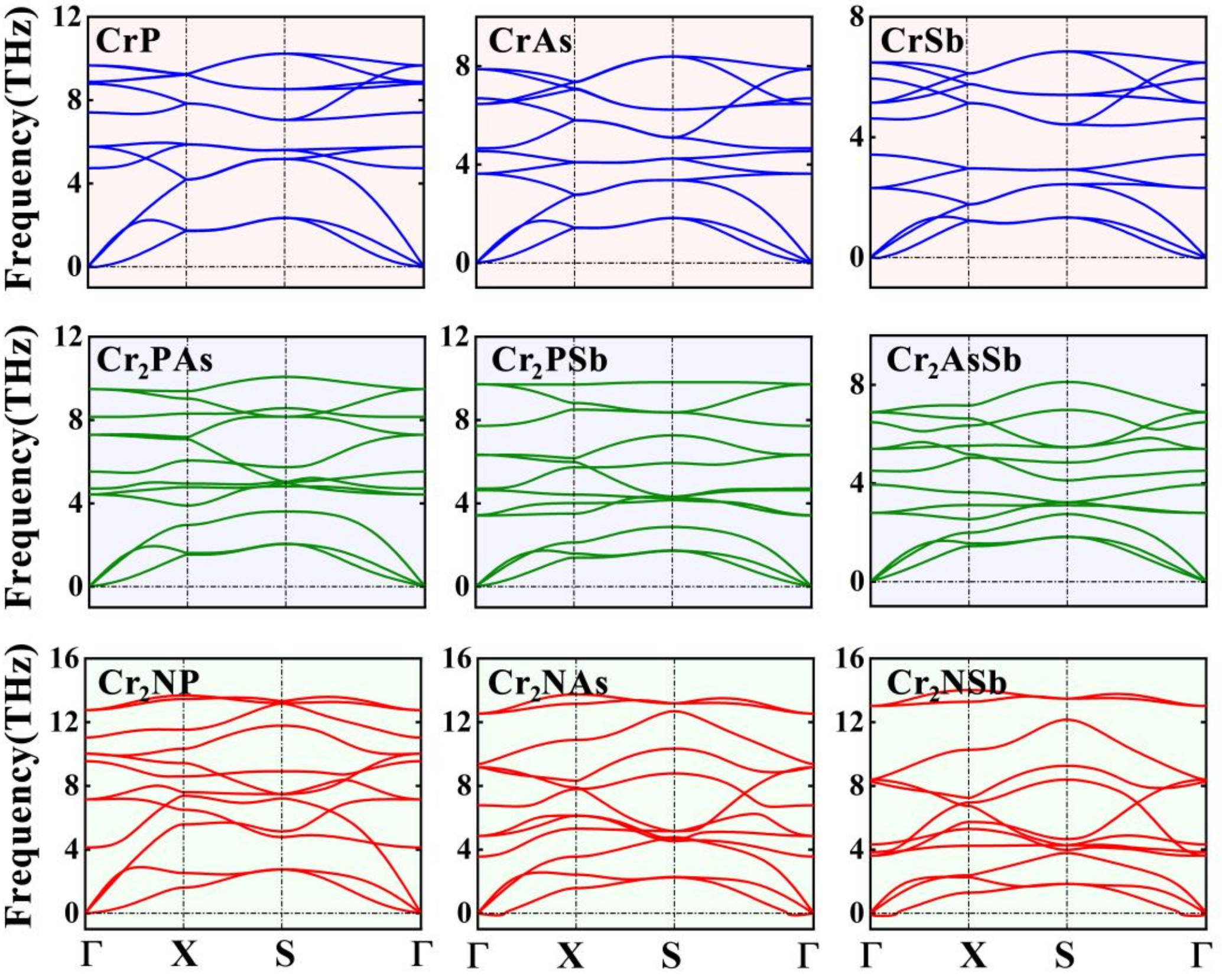

FIG. S1. Phonon dispersion of the CrY, $Cr_2XY$ and $Cr_2NY$ monolayers.

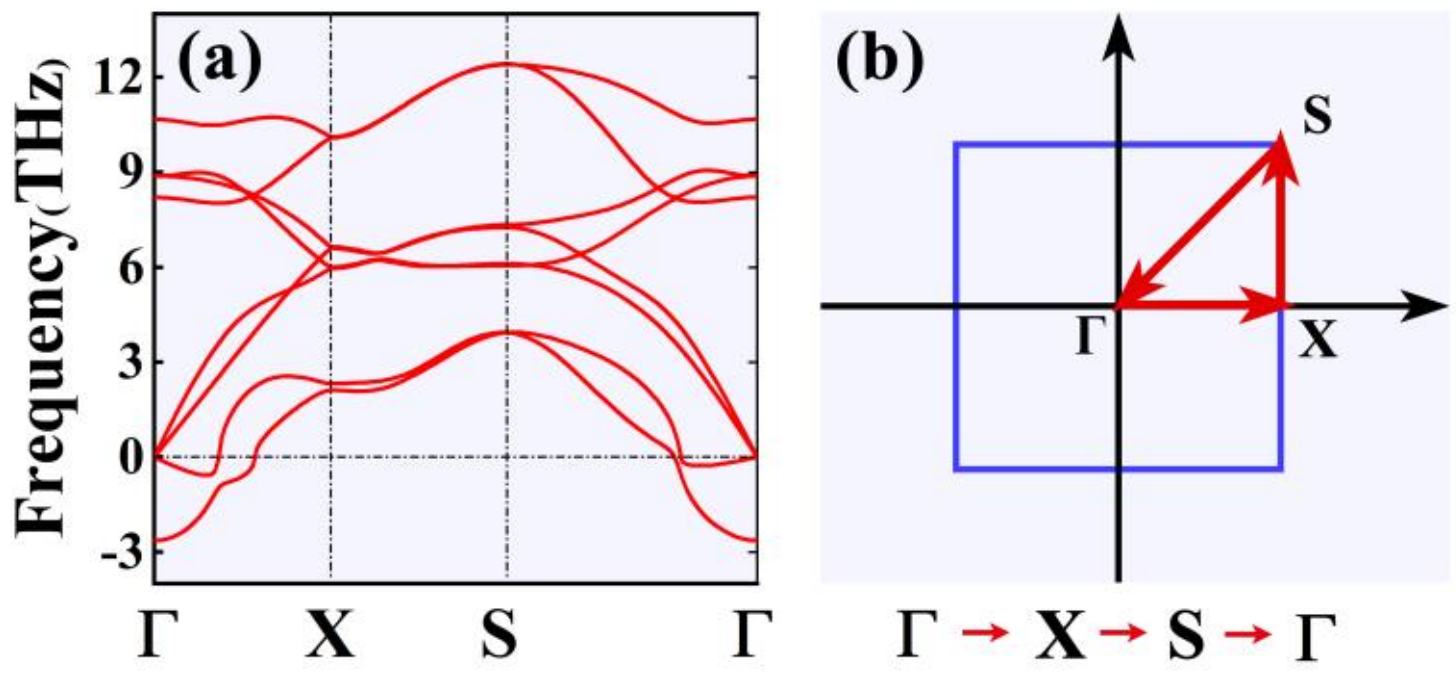

FIG. S2. (a) Phonon dispersion of the CrN monolayer.

(b) High symmetry *k*-path in the Brillouin zones.

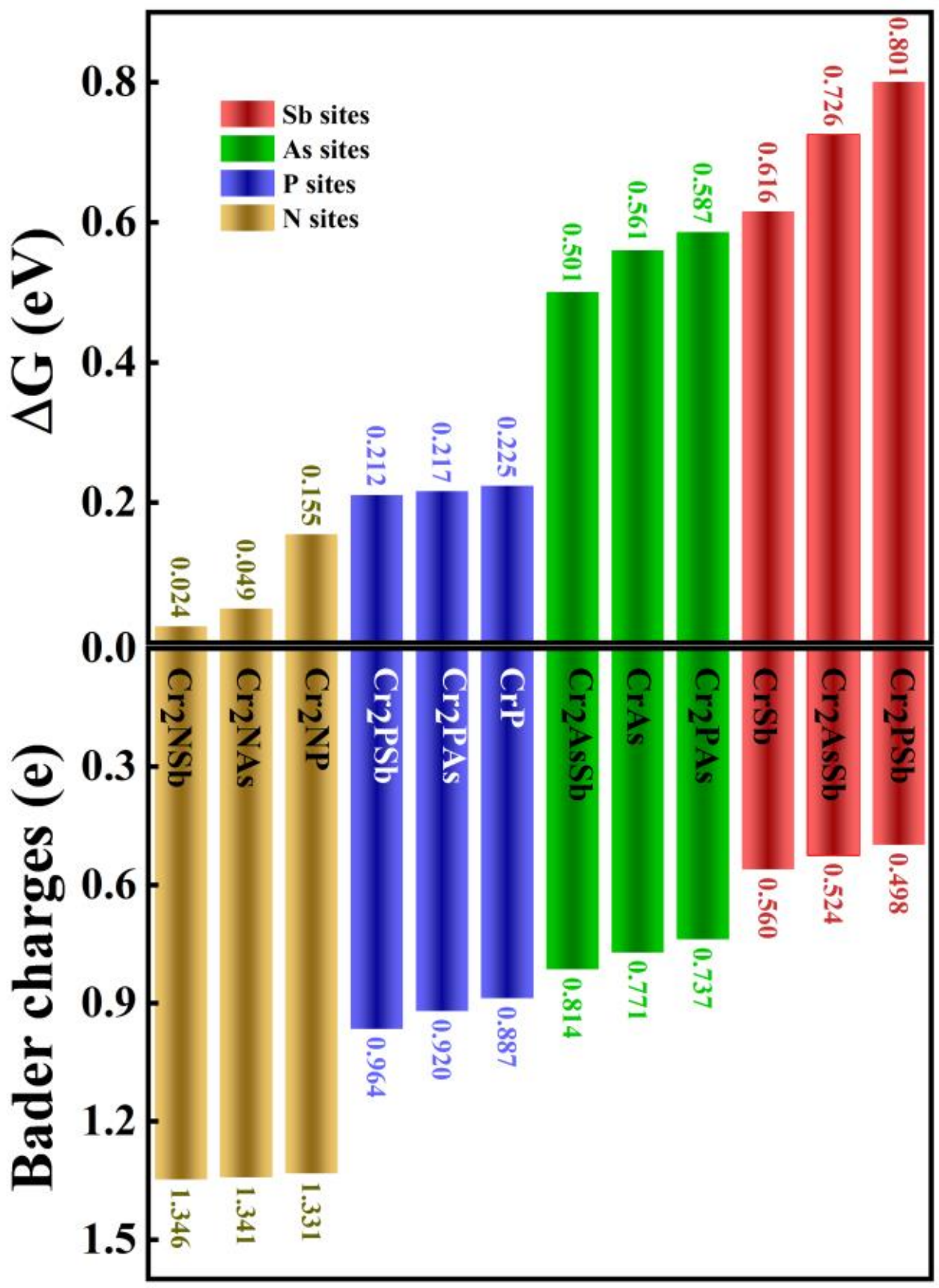

FIG. S3. Gibbs Free energy differences of adsorbing a hydrogen atom (top panel) and Bader charges (lower panel) at P, As, Sb sites of the CrX and $Cr_2XY$ monolayers and N sites of $Cr_2NY$ monolayers.

TABLE SI. The formation energies ($E_f$) and elastic constants ($C_{11}$, $C_{12}$ and $C_{33}$) of CrX, $Cr_2XY$ and $Cr_2NY$ monolayers.

| Monolayer | $E_f$ (eV) | $C_{11}$(N/m) | $C_{12}$(N/m) | $C_{66}$(N/m) |
|---|---|---|---|---|
| CrP | -2.86 | 36.46 | 16.96 | 28.29 |
| CrAs | -2.47 | 28.29 | 15.49 | 20.77 |
| CrSb | -1.74 | 16.68 | 6.44 | 11.46 |
| $Cr_2PAs$ | -2.67 | 34.24 | 15.17 | 24.05 |
| $Cr_2PSb$ | -2.34 | 32.25 | 12.57 | 19.23 |
| $Cr_2AsSb$ | -2.11 | 12.59 | 13.20 | 15.08 |
| $Cr_2NP$ | -3.03 | 75.20 | 33.79 | 48.48 |
| $Cr_2NAs$ | -2.52 | 72.57 | 29.51 | 42.40 |
| $Cr_2NSb$ | -2.21 | 71.17 | 35.02 | 36.42 |

TABLE SII. The Bader charges ($\Delta q_i$), atomic layer spacing ($\Delta z_i$) and local dipoles ($\mu_i$) along z direction between Cr and one of both sides of CrX, $Cr_2XY$ and $Cr_2NY$ monolayers.

| Monolayer | $\Delta q_1$(e) | $\Delta q_2$(e) | $\Delta z_1$(Å) | $\Delta z_2$(Å) | $\mu_1$(D) | $\mu_2$(D) |
|---|---|---|---|---|---|---|
| CrP | 0.89 | 0.89 | 1.20 | 1.20 | 1.07 | 1.07 |
| CrAs | 0.78 | 0.78 | 1.36 | 1.36 | 1.06 | 1.06 |
| CrSb | 0.57 | 0.57 | 1.66 | 1.66 | 0.95 | 0.95 |
| $Cr_2PAs$ | 0.92 | 0.74 | 1.16 | 1.40 | 1.07 | 1.04 |
| $Cr_2PSb$ | 0.97 | 0.50 | 1.08 | 1.73 | 1.05 | 0.87 |
| $Cr_2AsSb$ | 0.82 | 0.53 | 1.30 | 1.71 | 1.07 | 0.91 |
| $Cr_2NP$ | 1.33 | 0.75 | 0.52 | 1.43 | 0.69 | 1.07 |
| $Cr_2NAs$ | 1.34 | 0.60 | 0.46 | 1.60 | 0.62 | 0.96 |
| $Cr_2NSb$ | 1.35 | 0.33 | 0.37 | 1.93 | 0.50 | 0.64 |

***Note:*** The local dipoles ($\mu_i$) are only used to show the root of the larger dipole with abnormal direction for $Cr_2NY$ and have no real physical meaning.

TABLE SIII. The total amount of charge received by every atom ($\Delta\rho$) and dipole moment ($\mu$) of functionalized $Cr_2NP$.

| Monolayer | $\Delta\rho$(Cr) | $\Delta\rho$(N) | $\Delta\rho$(P) | $\Delta\rho$(H/F) | $\mu$(D) |
|---|---|---|---|---|---|
| $Cr_2NP$ | -2.08 | 1.33 | 0.75 | — | 0.39 |
| $Cr_2NPH$ | -2.00 | 1.32 | 0.32 | 0.362 | 0.21 |
| $Cr_2NHP$ | -1.97 | 1.43 | 0.89 | -0.350 | 1.24 |
| $Cr_2NPF$ | -2.02 | 1.31 | -0.05 | 0.763 | 0.68 |
| $Cr_2NHPH$ | -1.96 | 1.45 | 0.50 | -0.32/0.33 | 0.50 |
| $Cr_2NHPF$ | -1.84 | 1.39 | 0.06 | -0.35/0.74 | 1.53 |